\documentclass[fleqn,preprint,showpacs,preprintnumbers,amsmath,amssymb]{revtex4}
\usepackage{graphicx}% Include figure files
\usepackage{dcolumn}% Align table columns on decimal point
\usepackage{bm}% bold math
\begin{document}

\title{Propagation and oblique collision of electron-acoustic solitons in two-electron-populated quantum plasmas}
\author{M. Akbari-Moghanjoughi$^1$ and N. Ahmadzadeh-Khosroshahi$^2$}
\affiliation{$^1$ Azarbaijan University of
Tarbiat Moallem, Faculty of Science,
Department of physics, 51745-406, Tabriz, Iran\\
$^2$ Islamic Azad University, Hadishahr Branch, Hadishahr, Iran}

\date{\today \hspace{2mm}}
\begin{abstract}
Oblique interaction of small- but finite-amplitude KdV-type electron-acoustic solitary excitations is examined in an unmagnetized two-electron-populated degenerate quantum electron-ion plasma in the framework of quantum hydrodynamics model using the extended Poincar\'{e}-Lighthill-Kuo (PLK) perturbation method. Critical plasma parameter is found to distinguish the types of solitons and their interaction phase-shifts. It is shown that, depending on the critical quantum diffraction parameter $H_{cr}$, both compressive and rarefactive solitary excitations may exist in this plasma and their collision phase-shifts can be either positive or negative for the whole range of the collision angle $0<\theta<\pi$.
\end{abstract}


\maketitle

\section{Introduction}

There has been early reports on the existence of ordinary plasmas populated with two different electron species, namely, two-temperature-electron (2Te) plasmas \cite{bezzerides, mozer}. One of distinct nonlinear features of 2Te plasmas is their high frequency electron-acoustic (EA) spectrum compared to that of ion acoustic waves (IAWs). Properties of EA and ion-acoustic (IA) wave propagation and their modulational instability in ordinary 2Te plasmas has been extensively studied by many authors \cite{dell, gary, dubouloz, lakhina, goswami, esfand}. Electron-acoustic solitary waves (EASWs) have also been recently studied in unmagnetized \cite{sah} and magnetized \cite{hai} quantum plasmas. It has been found that, the characteristic features of the EASWs are significantly affected by the variation in the ratio of hot-to-cold (degenerate-to-nondegenerate) electron concentration \cite{misra} in planar as well as non-planar geometries. Recent investigation of electron-acoustic solitary propagations has shown that critical hot-to-cold electron densities may exist in two electron species quantum plasmas \cite{sahu}. It is also noted that the propagation of the electron-acoustic waves remain undamped in the range $0.25<n_{c0}/n_{h0}<4$ for ordinary 2Te plasmas\cite{tokar}, where, $n_{c0}$ and $n_{h0}$ denote the cold and hot electrons equilibrium densities, respectively. However, in a quantum plasma the degeneracy of one electron specie is characterized by higher relative number-density of that specie, meaning that in quantum plasmas $n_{c0}/n_{h0}\ll 1$.

Recently, quantum plasmas has attracted many attention and is becoming one of intense fields of plasma research \cite{haas, misra1, masood2, sabry, misra2} due to its broad application in manufacturing the micro- and nano-structured electronic devices \cite{Haug}. Quantum plasma is characterized by high densities and low temperatures contrary to ordinary plasma. The quantum hydrodynamics (QHD) model has also been applied by C. L. Gardner \cite{Gardner} to study the electron-hole dynamics in semiconductors. The quantum effects emerge when the de Broglie thermal-wavelength $\lambda_D = h/(2\pi m_e k_B T)^{1/2}$ of electrons become comparable to the inter-particle distances \cite{Bonitz}, the condition which is well satisfied for metallic and semiconductor compounds. The state of quantum plasma where the electron thermal energy is much less than their Fermi-energy is governed by the Fermi statistics and is the so called degenerate state. In a normal metal at room temperature the conduction electrons are mostly in degenerate state. Such conditions can also be found in the laboratory laser-produced plasmas \cite{Kremp, Andreev, Bingham}. The propagation of IASWs in quantum 2Te plasma has been recently studied in Refs. \cite{mahmood, masood1}, however, up to author's current knowledge there is no investigations of solitary wave interactions in such plasmas.

As it was first discovered by Zabysky et.al \cite{zabusky}, a unique feature of a solitary interaction is the asymptotic preservation of solitary wave amplitude. However, an imprint of solitary collisions can be a phase-shift in the post-collision trajectory of each wave. Two particular cases of solitary collisions are known as head-on and overtaking collisions. The former case has been extensively investigated using extended Poincar\'{e}-Lighthill-Kuo (PLK) perturbation method in one dimensional ordinary \cite{han, jiu} and quantum \cite{Labany} plasmas. On the other hand, the later case, in which the solitary waves propagate in the same direction, is studied using the inverse scattering transformation method \cite{gardner2}. A more general case is a planar collision under an arbitrary angle which requires two or three-dimensional \cite{Gui} treatment. The two-dimensional interaction of KdV-type solitons has been carried out in electron-ion (e-i) and dusty plasmas in non-relativistic \cite{jiu2, akbari, jiang} as well as weakly relativistic \cite{jiu3} cases.

In current work we aim at investigating the effect of relative electron degeneracy population on propagation and interaction of small-amplitude electron-acoustic solitary waves in a quantum two-electron-populated plasma. Although, the focus is on two-electron-populated (partially degenerated) quantum plasmas, however, it is easily noticed that current approach can be extended to the two-ion-populated quantum plasmas. The organization of the article is as follows. Description of quantum hydrodynamics state equations is given in Section 2. Reductive perturbation method is applied and the KdV evolution equation as well as phases are obtained in Section 3. Section 4 presents the discussions based on numerical analysis and, finally, Section 5 devotes to the concluding remarks.

\section{QHD Description of plasma}\label{equations}

Consider a three component dense plasma in which two types of inertial-less electrons exist in the presence of background inertial positive heavy ions. Also consider that one type of electrons are degenerate and others are non-degenerate. The classical pressure can be ignored for inertial heavy ions compared to much higher degeneracy pressure of electrons. The two-dimensional quantum continuum-pressure continuity equation closed set, which includes quantum tunneling effect for electrons \cite{Gardner}, can be written as
\begin{equation}\label{dimensional}
\begin{array}{l}
\frac{{\partial {n_d}}}{{\partial t}} + \nabla  \cdot {n_d}{{\vec V}_d} = 0, \\
\frac{{\partial {n_n}}}{{\partial t}} + \nabla  \cdot {n_n}{{\vec V}_n} = 0, \\
\frac{{\partial {u_d}}}{{\partial t}} + ({{\vec V}_d} \cdot \nabla) {{\vec V}_d} =  \frac{e}{{{m_e}}}\nabla \varphi  - \frac{1}{{{m_e}{n_d}}}\nabla {P_d} + \frac{{{\hbar ^2}}}{{2{m_e}^2}}\nabla \left[ {\frac{{{\nabla ^2}\sqrt {{n_d}} }}{{\sqrt {{n_d}} }}} \right], \\
\frac{{\partial {u_n}}}{{\partial t}} + ({{\vec V}_n} \cdot \nabla) {{\vec V}_n} =  \frac{e}{{{m_e}}}\nabla \varphi  + \frac{{{\hbar ^2}}}{{2{m_e}^2}}\nabla \left[ {\frac{{{\nabla ^2}\sqrt {{n_n}} }}{{\sqrt {{n_n}} }}} \right], \\
{\nabla ^2}\varphi  = \frac{e}{{{\varepsilon _0}}}({n_d} + {n_n} - {z_i}{n_i}), \\
\end{array}
\end{equation}
where, the $n$ and $d$ subscripts are used to label non-degenerate and degenerate species, respectively. The parameters $\hbar$, $N$ and $Z$ indicate the scaled Plank constant, background ion-density and atomic-number. Also, the quantities ${\vec V}_n$ (${\vec V}_d$), $n_n$ ($n_d$) and $\varphi$, refer to the velocity and number-density of \emph{non-degenerate} (\emph{degenerate}) electrons and the electrostatic potential, respectively. The normalized parameter $H=\hbar \omega_{pd}/2k_B T_{Fd}$ is the quantum diffraction parameter, which is the ratio of (d)-electron plasmon-energy to (d)-electron Fermi-energy. It is noted that, in a dense and degenerate plasma environment and under the zero-temperature Fermi-gas assumption, the degeneracy pressure of (d)-electrons obeys the Pauli exclusion principle and is related to the equilibrium (d)-electron number-density through the following relations in two dimension \cite{Shukla0}
\begin{equation}
P = \frac{{{m_e}v_{Fe}^2{n^{(0)}}}}{4}{\left( {\frac{n}{{{n^{(0)}}}}} \right)^2},{v_{Fe}} = \sqrt {\frac{{2{E_{Fe}}}}{{{m_e}}}} ,{E_{Fe}} = {k_B}{T_{Fe}},
\end{equation}
where, quantities $v_{Fe}$, $E_{Fe}$ and $T_{Fe}$ are the Fermi-velocity, Fermi-energy and Fermi-temperature of degenerate-electrons, respectively. The quantity $n^{(0)}$ denotes the electrons equilibrium number-density. From the standard definitions, it is also known that in a \emph{two-dimensional} degenerate Fermi-gas the quantum equilibrium number-density is related to the Fermi-temperature of degenerate electrons, i.e. \cite{Landau}
\begin{equation}\label{T}
{T_{Fe}} = \frac{{{\hbar ^2}}}{{2{m_e}}}\left( {\frac{{2{\pi}{n^{(0)}}}}{{{k_B}}}} \right).
\end{equation}
From Eq. (\ref{T}), it is observed that even in $H\rightarrow 0$ limit the classical model is not retained. The normalized equations may be obtained using the following re-scalings from basic equations (Eqs. (\ref{dimensional}))
\begin{equation}
x \to \frac{{{C_{sd }}}}{{{\omega _{pd }}}}x,t \to \frac{t}{{{\omega _{pd }}}},n \to nn_d^{(0)},{\vec V} \to {\vec V}{v_{Fe}},\varphi  \to \varphi \frac{{2{k_B}{T_{Fd}}}}{e}.
\end{equation}
where, ${\omega _{pd}} = \sqrt {{e^2}n_{d}^{(0)}/{\varepsilon _0}{m_e}}$ and ${v_{Fe}} = \sqrt {2{k_B}{T_{Fd}}/{m_e}}$ are the characteristic plasma-frequency and Fermi-speed, respectively. The normalized set of closed QHD equations, therefore, read as
\begin{equation}\label{normal}
\begin{array}{l}
\frac{{\partial {n_\alpha }}}{{\partial t}} + \nabla \cdot{n_\alpha }{{\vec V}_\alpha } = 0,{{\vec V}_\alpha } = \mathord{\buildrel{\lower3pt\hbox{$\scriptscriptstyle\frown$}}
\over i} {u_\alpha } + \mathord{\buildrel{\lower3pt\hbox{$\scriptscriptstyle\frown$}}
\over j} {v_\alpha }, \\
\frac{{\partial {{\vec V}_\alpha }}}{{\partial t}} + ({{\vec V}_\alpha} \cdot \nabla){{\vec V}_\alpha} =  \nabla \varphi  - \frac{{{D_{\alpha} }}}{2}\nabla {n_\alpha } + \frac{{{H^2}}}{2}\nabla \left[ {\frac{{{\nabla ^2}\sqrt {{n_\alpha }} }}{{\sqrt {{n_\alpha }} }}} \right], \\
{\nabla ^2}\varphi  = \sum\limits_\alpha  {{n_\alpha }}  - {z_i}{n_i}. \\
\end{array}
\end{equation}
Here, the label $\alpha$ is used to denotes (d)/(n)-electrons, for simplicity. The new quantities, introduced here, are $D_{\alpha}=\{1,0\}$ and $n_{\alpha}^{(0)}=\{1,\beta\}$ for $\alpha=\{d,n\}$, respectively. On the other hand, the quasi-neutrality condition at thermodynamics equilibrium is given by the Poisson's relation as
\begin{equation}
n_{d}^{(0)} +n_{n}^{(0)} -z_i n_i =0,
\end{equation}
or in a reduced form
\begin{equation}\label{neural}
\delta =1+\beta,\hspace{3mm}\beta =\frac{n_{n}^{(0)}}{n_{d}^{(0)}},\hspace{3mm}\delta =\frac{z_i n_i }{n_{d}^{(0)} }.
\end{equation}
Considering two small perturbations which move at angle, $\theta$, in $x-y$ plane with different velocities will approach to each other, they can interact at some event and finally depart leaving a phase-shift on each others trajectories. To evaluate the dynamics of this collision, we use asymptotic expansion of plasma variables around thermodynamics equilibrium state in an appropriate strained coordinate which contains the phase records of each wave. The technique is so-called extended Poincar\'{e}-Lighthill-Kuo (PLK) method \cite{jeffery, masa}. The reductive perturbation method in nonlinear wave propagation has been put forward by Taniuti et al. \cite{tosiya}, where it is shown that the stretching of $\xi  = {\varepsilon ^a}(x - ct)$, $\tau  = {\varepsilon ^{a + 1}}$ and $a = {(p - 1)^{ - 1}}$ with $p=2$, $p=3$ admit reductions to Burgers and KdV equations, respectively. Hence, the normalized equation set are introduced to the following strained coordinate
\begin{equation}\label{stretch}
\begin{array}{l}
\xi = \varepsilon ({k_1}x + {l_1}y - {c_\xi}t) + {\varepsilon ^2}{P_0}(\eta,\tau) + {\varepsilon ^3}{P_1}(\xi,\eta,\tau) +\ldots,\\
\eta = \varepsilon ({k_2}x + {l_2}y - {c_\eta}t) + {\varepsilon ^2}{Q_0}(\xi,\tau) + {\varepsilon ^3}{Q_1}(\xi,\eta,\tau) + \ldots,\\
\tau = {\varepsilon ^3}t,\\c_\xi=c_1, \hspace{3mm}c_\eta=c_2.
\end{array}
\end{equation}
The functions $P_j$ and $Q_j$ ($j=0,1,2,...$) are the phase record of the interacting waves in space-time and are to be determined later along with the evolution equations in section \ref{derivation}. Initial wave trajectories are given by vectors $\textbf{r}_1=(k_1,l_1)$ and $\textbf{r}_2=(k_2,l_2)$ and the collision angle $\theta$ is defined by
\begin{equation}
\begin{array}{l}
\cos\theta = \frac{{{\mu}}}{{{\lambda _1}{\lambda _2}}},\\
\mu  = ({k_1}{k_2} + {l_1}{l_2}),\\
{\lambda _1} = {(k_1^2 + l_1^2)^{\frac{1}{2}}},\\
{\lambda _2} = {(k_2^2 + l_2^2)^{\frac{1}{2}}},
\end{array},
\end{equation}
where, $\lambda _1$ and $\lambda _2$ are the normalized wave-numbers. The asymptotic expansion of the dependent plasma variables in powers of $\varepsilon$ in the states away from thermodynamics equilibrium is carried out using the following orderings
\begin{equation}\label{Ordering}
\begin{array}{l}
\left( {\begin{array}{*{20}{c}}
{{n_\alpha }}  \\
\begin{array}{l}
{u_\alpha } \\
{v_\alpha } \\
\end{array}  \\
\varphi   \\
\end{array}} \right) = \left( {\begin{array}{*{20}{c}}
{n_\alpha ^{(0)}}  \\
\begin{array}{l}
0 \\
0 \\
\end{array}  \\
0  \\
\end{array}} \right) + {\varepsilon ^2}\left( {\begin{array}{*{20}{c}}
{n_\alpha ^{(1)}}  \\
\begin{array}{l}
u_\alpha ^{(1)} \\
v_\alpha ^{(1)} \\
\end{array}  \\
{{\varphi ^{(1)}}}  \\
\end{array}} \right) + {\varepsilon ^4}\left( {\begin{array}{*{20}{c}}
{n_\alpha ^{(2)}}  \\
\begin{array}{l}
u_\alpha ^{(2)} \\
v_\alpha ^{(2)} \\
\end{array}  \\
{{\varphi ^{(2)}}}  \\
\end{array}} \right) +  \cdots
\end{array}
\end{equation}
where, the parameter $\varepsilon$ is a very small, positive and real value proportional to the perturbation amplitude. \textbf{Note also that, we use the stretching employed in Ref. \cite{jiu3}, which is completely different from that in the cases of two- and three-dimensional magnetized \cite{gui, akbari2} and of unmagnetized one-dimensional collisions \cite{han}.} The reduced set of plasma equations in strained coordinates are given in appendix A, for simplicity. From the leading-order in $\varepsilon$ in Eqs. (\ref{strain1}-\ref{strain4}), we isolate the following relations
\begin{equation}\label{leading1}
\left( {{c_1}\frac{\partial }{{\partial \xi }} + {c_2}\frac{\partial }{{\partial \eta }}} \right)n_\alpha ^{(1)} = n_\alpha ^{(0)}\left( {{k_1}\frac{\partial }{{\partial \xi }} + {k_2}\frac{\partial }{{\partial \eta }}} \right)u_\alpha ^{(1)} + n_\alpha ^{(0)}\left( {{l_1}\frac{\partial }{{\partial \xi }} + {l_2}\frac{\partial }{{\partial \eta }}} \right)v_\alpha ^{(1)},
\end{equation}
\begin{equation}
\left( {{c_1}\frac{\partial }{{\partial \xi }} + {c_2}\frac{\partial }{{\partial \eta }}} \right)u_\alpha ^{(1)} = \frac{{D_{\alpha} }}{2}\left( {{k_1}\frac{\partial }{{\partial \xi }} + {k_2}\frac{\partial }{{\partial \eta }}} \right)n_\alpha ^{(1)} - \left( {{k_1}\frac{\partial }{{\partial \xi }} + {k_2}\frac{\partial }{{\partial \eta }}} \right){\varphi ^{(1)}},
\end{equation}
\begin{equation}
\left( {{c_1}\frac{\partial }{{\partial \xi }} + {c_2}\frac{\partial }{{\partial \eta }}} \right)v_\alpha ^{(1)} = \frac{{D_{\alpha} }}{2}\left( {{l_1}\frac{\partial }{{\partial \xi }} + {l_2}\frac{\partial }{{\partial \eta }}} \right)n_\alpha ^{(1)} - \left( {{l_1}\frac{\partial }{{\partial \xi }} + {l_2}\frac{\partial }{{\partial \eta }}} \right){\varphi ^{(1)}},
\end{equation}
\begin{equation}\label{leading4}
\sum\limits_\alpha  {n_\alpha ^{(1)} = 0} .  \\
\end{equation}
Consequently, we obtain the following first-order perturbed plasma components by solving the coupled equations (Eqs. (\ref{leading1}-\ref{leading4}))
\begin{equation}\label{Firstcomp}
{\left( {\begin{array}{*{20}{c}}
{n_\alpha ^{(1)}}  \\
{\begin{array}{*{20}{c}}
{u_\alpha ^{(1)}}  \\
{v_\alpha ^{(1)}}  \\
\end{array}}  \\
{{\varphi ^{(1)}}}  \\
\end{array}} \right) = \left( {\begin{array}{*{20}{c}}
{{A_{1\alpha }}}  \\
{\begin{array}{*{20}{c}}
{{A_{2\alpha }}}  \\
{{A_{3\alpha }}}  \\
\end{array}}  \\
1  \\
\end{array}} \right){\varphi ^{(1)}}(\xi,\tau) + \left( {\begin{array}{*{20}{c}}
{{B_{1\alpha }}}  \\
{\begin{array}{*{20}{c}}
{{B_{2\alpha }}}  \\
{{B_{3\alpha }}}  \\
\end{array}}  \\
1  \\
\end{array}} \right){\varphi ^{(1)}}(\eta,\tau).}
\end{equation}
The quantities ${\varphi ^{(1)}}(\xi,\tau)$ and ${\varphi ^{(1)}}(\eta,\tau)$ describe the first-order amplitude evolution of two distinct solitary waves in oblique directions ${\eta_ \bot }$ and ${\xi_ \bot }$, respectively, as we will see in the next section. In the proceeding calculations we will use the notations ${\varphi_\xi ^{(1)}}$ and ${\varphi_\eta ^{(1)}}$ instead of ${\varphi ^{(1)}}(\xi,\tau)$ and ${\varphi ^{(1)}}(\eta,\tau)$, for brevity. The coefficients of first-order components of plasma variable approximations for (d)- and (n)-electrons (labeled by $\alpha$), are given as
\begin{equation}
\begin{array}{*{20}{c}}
{{A_{1\alpha }} = \frac{2{n_\alpha ^{(0)}\lambda _1^2}}{{n{_\alpha ^{(0)}}{D_{\alpha}}\lambda _1^2-2c_1^2}},\vspace{3mm}{B_{1\alpha }} = \frac{2{n_\alpha ^{(0)}\lambda _2^2}}{{n{{_\alpha ^{(0)}}}{D_{\alpha}}\lambda _2^2-2c_2^2}},}  \\
{{A_{2\alpha }} = \frac{2{{c_1}{k_1}}}{{n{{_\alpha ^{(0)}}}{D_{\alpha}}\lambda _1^2-2c_1^2}},\vspace{3mm}{B_{2\alpha }} = \frac{2{{c_2}{k_2}}}{{n{{_\alpha ^{(0)}}}{D_{\alpha}}\lambda _2^2-2c_2^2}},}  \\
{{A_{3\alpha }} = \frac{2{{c_1}{l_1}}}{{n{{_\alpha ^{(0)}}}{D_{\alpha}}\lambda _1^2-2c_1^2}},\vspace{3mm}{B_{3\alpha }} = \frac{2{{c_2}{l_2}}}{{n{{_\alpha ^{(0)}}}{D_{\alpha}}\lambda _2^2-2c_2^2}}.}  \\
\end{array}
\end{equation}
The nonlinear dispersion relations, which is deduced from the first-order approximations, can be written in the following compact form
\begin{equation}\label{dispers}
\sum\limits_\alpha  {\frac{{n_\alpha ^{(0)}}}{{2c_1^2 - D_{\alpha}n{{_\alpha ^{(0)}}}\lambda _1^2}}}  = \sum\limits_\alpha  {\frac{{n_\alpha ^{(0)}}}{{2c_2^2 - D_{\alpha}n{{_\alpha ^{(0)}}}\lambda _2^2}}}  = 0.
\end{equation}
Also, the normalized phase-speeds $c_1$ and $c_2$ of waves are given as
\begin{equation}\label{speed}
{c_1} = {\lambda _1}\sqrt {\frac{\beta }{2({1 + \beta })}} ,{c_2} = {\lambda _2}\sqrt {\frac{\beta }{2({1 + \beta })}} .
\end{equation}

\section{Solitary Collision Dynamics}\label{derivation}

In this section from the second-order, we derive the nonlinear wave evolution equations and the related phase-functions introduced in Eqs. (\ref{stretch}). We use Eqs. (\ref{strain1}-\ref{strain4}) to deduce relations among higher-order (second-order in $\varepsilon$) plasma variables, and consequently, by solving the coupled differential equations in this approximation level and by using the dispersion relations (Eq. (\ref{dispers})), the second-order d/n-electrons number-density perturbation components are obtained in the following form
\begin{equation}{\label{n2}}
\begin{array}{l}
n_d ^{(2)} = K_1 N_1\left[ {\frac{{\partial \varphi _\xi^{(1)}}}{{\partial \tau}} + {A_1}\varphi _\xi^{(1)}\frac{{\partial \varphi _\xi^{(1)}}}{{\partial \xi}} + {B_1}\frac{{{\partial ^3}\varphi _\xi^{(1)}}}{{\partial {\xi^3}}}} \right]\eta +  \\
K_2 N_2\left[ {\frac{{\partial \varphi _\eta^{(1)}}}{{\partial \tau}} + {A_2}\varphi _\eta^{(1)}\frac{{\partial \varphi _\eta^{(1)}}}{{\partial \eta}} + {B_2}\frac{{{\partial ^3}\varphi _\eta^{(1)}}}{{\partial {\eta^3}}}} \right]\xi +  \\
{K_1{E_2}}\left[ {{P_0}(\eta,\tau) - \frac{E_1}{E_2}\smallint \varphi _\eta^{(1)}d\eta} \right]\frac{{\partial \varphi _\xi^{(1)}}}{{\partial \xi}} +  \\
{K_2{E'_2}}\left[ {{Q_0}(\xi,\tau) - \frac{E'_1}{E'_2}\smallint \varphi _\xi^{(1)}d\xi} \right]\frac{{\partial \varphi _\eta^{(1)}}}{{\partial \eta}} +  \\
\left[ {\frac{{K_1}{C_1}}{2}{{(\varphi _\xi^{(1)})}^2} + ({K_1}D_1 + {K_2}D_2)\varphi _\xi^{(1)}\varphi _\eta^{(1)} + \frac{{K_2}{C_2}}{2}{{(\varphi _\eta^{(1)})}^2}} \right] + \\
\left[{ {K_1}N_1\smallint \frac{{\partial \varphi _\xi^{(1)}}}{{\partial \tau}}d\xi +{K_2} N_2\smallint \frac{{\partial \varphi _\eta^{(1)}}}{{\partial \tau}}d\eta} \right] -  \\
\left[ {({L}\lambda _1^2 + {K_1 R_1})\frac{{{\partial ^2}\varphi _\xi ^{(1)}}}{{\partial {\xi ^2}}} + ({L}\lambda _2^2 + {K_2 R_2})\frac{{{\partial ^2}\varphi _\eta ^{(1)}}}{{\partial {\eta ^2}}}} \right] +\\ F(\xi ,\tau ) + G(\eta ,\tau ), \\
n_n ^{(2)} = \left[ {\lambda _1^2\frac{{{\partial ^2}\varphi _\xi ^{(1)}}}{{\partial {\xi ^2}}} + \lambda _2^2\frac{{{\partial ^2}\varphi _\eta ^{(1)}}}{{\partial {\eta ^2}}}} \right] - n_d^{(2)},
\end{array}
\end{equation}
where, the functions $F(\xi,\tau)$ and $G(\eta,\tau)$ represent the homogenous solutions of integrals involved in $n_\alpha ^{(2)}$ differential equations. The unknown coefficients appearing in Eq. (\ref{n2}) are given below
\begin{equation}\label{coeffs1}
\begin{array}{l}
\frac{{{A_1}}}{{{\lambda _1}}} = \frac{{{A_2}}}{{{\lambda _2}}} =  - \frac{{(1 - \beta )(3 + 2\beta )}}{{1 + 3\beta }}\sqrt {\frac{{2(1 + \beta )}}{\beta }},
\end{array}
\end{equation}
\begin{equation}
\begin{array}{l}
\frac{{{B_1}}}{{{\lambda _1}^3}} = \frac{{{B_2}}}{{{\lambda _2}^3}} = \frac{{{H^2}{{(1 + \beta )}^3} - \beta }}{{2\sqrt 2 \sqrt \beta  {{(1 + \beta )}^{3/2}}(1 + 3\beta )}},
\end{array}
\end{equation}
\begin{equation}
\begin{array}{l}
{E_1} ={E'_1}= \frac{{4{{(1 + \beta )}^2}(3 + \beta (8 + \beta ) - (1 - \beta )\beta \cos\theta)}}{\beta },
\end{array}
\end{equation}
\begin{equation}
\begin{array}{l}
\frac{{E_2}{\lambda _1}}{{\lambda _2}} = \frac{{E'_2}{\lambda _2}}{{\lambda _1}} = 2(1 + \beta )(1 + 3\beta )(1 - \cos\theta),
\end{array}
\end{equation}
\begin{equation}
\begin{array}{l}
\frac{{C_1}{\lambda _2}}{{\lambda _1}} = \frac{{C_2}{\lambda _1}}{{\lambda _2}} = \frac{{4{{(1 + \beta )}^2}(2\beta (\beta  - 1) - (1 + {\beta ^2})\cos\theta)}}{\beta },\\
\end{array}
\end{equation}
\begin{equation}
\begin{array}{l}
\frac{{{D_1}{\lambda _2}}}{{{\lambda _1}}} = \frac{{{D_2}{\lambda _1}}}{{{\lambda _2}}} =  - \frac{{4{{(1 + \beta )}^2}(1 - \beta )(\beta  + \cos \theta )}}{\beta },\\
\end{array}
\end{equation}
\begin{equation}
\begin{array}{l}
\frac{{K_1}{{\lambda _2}}}{{\lambda _1}} = \frac{{K_2}{{\lambda _1}}}{{\lambda _2}} =\frac{L}{2}=\frac{1}{{2(1 + \beta )(1 - \cos\theta)}},
\end{array}
\end{equation}
\begin{equation}
\begin{array}{l}
{N_1}{\lambda _1} = {N_2}{\lambda _2} = \frac{{2\sqrt 2 (1 + 3\beta ){{(1 + \beta )}^{3/2}}}}{{\sqrt \beta  }},
\end{array}
\end{equation}
\begin{equation}
\begin{array}{l}
\frac{{{R_1}}}{{\lambda _1^2}} = \frac{{{R_2}}}{{\lambda _2^2}} = -\frac{{{H^2}{{(1 + \beta )}^3}}}{\beta},
\end{array}
\end{equation}
In Eq. (\ref{n2}), it is noted that, the first and the second terms diverge as $\xi\rightarrow\pm\infty$ and $\eta\rightarrow\pm\infty$, thus, their coefficients must vanish in order to eliminate the secularities. This leads to pair of distinct KdV-type evolution equations which describe the propagations of two solitary structures moving at $\xi\bot$ and $\eta\bot$ directions. The next two terms in Eq. (\ref{n2}) although are not secular at this order but will become ones in the next-order. It has been shown \cite{nob} that, making use of the method of multiple scales combined with the reductive perturbation technique, not only eliminates the secularities arising in the second order correction but also the phase factor of the lowest KdV soliton suffers a modification proportional to its amplitude. A fuller discussion of the elimination of secularities in higher-order amplitude approximation is given elsewhere \cite{masa}. From the later requirement the phase-shifts introduced in Eqs. (\ref{stretch}) are derived. Therefore, considering the mentioned requirements the low-amplitude dynamics (propagation and collision) of both solitary wave are defined by the following two coupled set of equations
\begin{equation}\label{kdv1}
\begin{array}{l}
\frac{{\partial \varphi _\xi^{(1)}}}{{\partial \tau}} + A_1\varphi _\xi^{(1)}\frac{{\partial \varphi _\xi^{(1)}}}{{\partial \xi}} + B_1\frac{{{\partial ^3}\varphi _\xi^{(1)}}}{{\partial {\xi^3}}} = 0,
\end{array}
\end{equation}
\begin{equation}\label{P0}
\begin{array}{l}
{P_0}(\eta,\tau) = \frac{{{E_1}}}{{{E_2}}}\int {\varphi _\eta^{(1)}d\eta},
\end{array}
\end{equation}
\begin{equation}\label{kdv2}
\begin{array}{l}
\frac{{\partial \varphi _\eta^{(1)}}}{{\partial \tau}} + A_2\varphi _\eta^{(1)}\frac{{\partial \varphi _\eta^{(1)}}}{{\partial \eta}} + B_2\frac{{{\partial ^3}\varphi _\eta^{(1)}}}{{\partial {\eta^3}}} = 0,
\end{array}
\end{equation}
\begin{equation}\label{Q0}
\begin{array}{l}
{Q_0}(\xi,\tau) = \frac{{{E'_1}}}{{{E'_2}}}\int {\varphi _\xi^{(1)}d\xi},
\end{array}
\end{equation}
Therefore, after elimination of secularities the second-order d-electrons number-density read as
\begin{equation}
\begin{array}{l}
n_d ^{(2)} = \left[ {{K_1}\left( {\frac{{{C_1}}}{2} - {A_1}{N_1}} \right)\varphi _\xi ^{{{(1)}^2}} + {K_2}\left( {\frac{{{C_2}}}{2} - {A_2}{N_2}} \right)\varphi _\eta ^{{{(1)}^2}}} \right] -  \\
({K_1}{N_1}{B_1} + {L}\lambda _1^2+K_1 R_1)\frac{{{\partial ^2}\varphi _\xi ^{(1)}}}{{\partial {\xi ^2}}} - ({K_2}{N_2}{B_2} + {L}\lambda _2^2+K_2 R_2)\frac{{{\partial ^2}\varphi _\eta ^{(1)}}}{{\partial {\eta ^2}}} +  \\
({K_1}{D_1} + {K_2}{D_2})\varphi _\xi ^{(1)}\varphi _\eta ^{(1)} + F(\xi ,\tau ) + G(\eta ,\tau ), \\
\end{array}
\end{equation}
There are multi-soliton solutions for (Eq. (\ref{kdv1}) and (\ref{kdv2})), however, a unique stationary single-soliton solutions are obtained by applying the appropriate boundary conditions, which require the perturbed potential components and their derivatives to vanish at infinities, i.e.
\begin{equation}\label{boundary}
\begin{array}{l}
\mathop {\lim }\limits_{\zeta \to \pm\infty } \{\varphi _\zeta^{(1)},\frac{\partial \varphi _\zeta^{(1)}}{\partial \zeta },\frac{\partial ^2\varphi _\zeta^{(1)}}{\partial \zeta
^2}\}=0,\hspace{3mm} \zeta={\xi,\eta},
\end{array}
\end{equation}
These suitable solutions are given as
\begin{equation}\label{phi-x}
\begin{array}{l}
{\varphi _\xi ^{(1)} = \frac{{{\varphi _{\xi 0}}}}{{\cosh^2(\frac{{\xi  - {u_{\xi 0}}\tau }}{{{\Delta _\xi }}})}},}  \\
{{\varphi _{\xi 0}} = \frac{{3{u_{\xi 0}}}}{{{A_1}}},{\Delta _\xi } = {{(\frac{{4{B_1}}}{{{u_{\xi 0}}}})}^{\frac{1}{2}}},}  \\
\end{array}
\end{equation}
\begin{equation}\label{phi-y}
\begin{array}{l}
\varphi _\eta^{(1)} = \frac{{{\varphi _{\eta0}}}}{{\cosh^2(\frac{{\eta - {u_{\eta0}}\tau}}{{{\Delta _\eta}}})}},\\
{\varphi _{\eta0}} = \frac{{3{u_{\eta0}}}}{A_2},\hspace{3mm} {\Delta _\eta} = {(\frac{{4B_2}}{{{u_{\eta0}}}})^{\frac{1}{2}}}.
\end{array}
\end{equation}
where, $\varphi _{\xi0}$ ($\varphi _{\eta0}$) and $\Delta_{\xi}$ ($\Delta_{\eta}$) represent the soliton amplitude and width, respectively, and $u_{\xi0}$ ($u_{\eta0}$) is an arbitrary value for relative soliton speed.

The first-order approximations for the collisional phase-shifts due to elastic collision of solitary excitations are obtained from Eqs. (\ref{P0}) and (\ref{Q0}) with use of KdV solutions (Eqs. (\ref{phi-x}) and (\ref{phi-y}))
\begin{equation}\label{phase-x}
\begin{array}{l}
{P_0}(\eta,\tau) = \frac{E_1}{E_2} \varphi _{\eta0} \Delta _\eta\tanh(\frac{{\eta - {u_{\eta0}}\tau}}{{{\Delta _\eta}}}),
\end{array}
\end{equation}
\begin{equation}\label{phase-y}
\begin{array}{l}
{Q_0}(\xi,\tau) = \frac{E'_1}{E'_2} \varphi _{\xi0} \Delta _\xi\tanh(\frac{{\xi - {u_{\xi0}}\tau}}{{{\Delta _\xi}}}).
\end{array}
\end{equation}
Therefore, the trajectories and the phase variations of the two collided solitary excitations up to order $O(\varepsilon^2)$ are fully determined by the following oblique coordinate
\begin{equation}\label{trajectory}
\begin{array}{l}
\xi = \varepsilon ({k_1}x + {l_1}y - {c_1}t) - \varepsilon^2\frac{{{E_1}}}{{{E_2}}}{\varphi _{\eta0}}{\Delta _\eta}\tanh (\frac{{\eta - {u_{\eta0}}\tau}}{{{\Delta _\eta}}})+O(\varepsilon^3), \\
\eta = \varepsilon ({k_2}x + {l_2}y - {c_2}t)] - \varepsilon^2\frac{{{{E'}_1}}}{{{{E'}_2}}}{\varphi _{\xi0}}{\Delta _\xi}\tanh (\frac{{\xi - {u_{\xi0}}\tau}}{{{\Delta _\xi}}})+O(\varepsilon^3). \\
\end{array}
\end{equation}
The overall phase-shifts, however, are obtained by comparing the phases of each wave long before and after the collision. In other words, the past-collision state of soliton labeled $"2"$ moving in ${\xi_ \bot }$ direction, described from the rest frame $\eta=0$ of soliton labeled $"1"$, is ($\xi=-\infty, \eta=0$) at a fixed time $t=const.$ and the post-collision state is ($\xi=+\infty, \eta=0$) at a fixed time ($t=const.$). On the other hand, the past-collision state of soliton labeled $"1"$ moving in ${\eta_ \bot }$ direction, described from the rest frame $\xi=0$ of soliton labeled $"2"$, is ($\xi=0, \eta=-\infty$) at a fixed time ($t=const.$) and the post-collision state is ($\xi=0, \eta=+\infty$) at a fixed time ($t=const.$). Thus, we can deduce the overall phase-shifts in the trajectories of each soliton in the following mathematical manner
\begin{equation}
\begin{array}{l}
\Delta {P_0} = P_{post-collision}-P_{past-collision}=\\ \mathop {\lim }\limits_{\xi=0,\eta \to  + \infty } [\varepsilon ({k_1}x + {l_1}y - {c_1}t)]-
\mathop {\lim }\limits_{\xi=0,\eta \to  - \infty } [\varepsilon ({k_1}x + {l_1}y - {c_1}t)] , \\
\Delta {Q_0} = Q_{post-collision}-Q_{past-collision}=\\ \mathop {\lim }\limits_{\eta=0,\xi \to  + \infty } [\varepsilon ({k_2}x + {l_2}y - {c_2}t)]-
\mathop {\lim }\limits_{\eta=0,\xi \to  - \infty } [\varepsilon ({k_2}x + {l_2}y - {c_2}t)] , \\
\end{array}
\end{equation}
where, the functions $\Delta {P_0}$ and $\Delta {Q_0}$ denote the overall phase-shifts of waves $"1"$ and $"2"$ in the oblique collision. By using Eqs. (\ref{phase-x}) and (\ref{phase-y}), we readily obtain
\begin{equation}\label{shifts}
\begin{array}{l}
\Delta {P_0} = 2{\varepsilon ^2}\frac{E_1}{E_2} \varphi _{\eta0} \Delta _\eta, \\
\Delta {Q_0} = 2{\varepsilon ^2}\frac{E'_1}{E'_2} \varphi _{\xi0} \Delta _\xi,
\end{array}
\end{equation}
The similar representations are given in Ref. \cite{jiu3}.

\section{Numerical Analysis and Discussion}\label{discussion}

Before the discussion a few comments concerning the possibility of solitary excitations in this plasma are in order. As it is evident from the KdV equation coefficients, the coefficients "$A$" vanish at the critical fractional degenerate-electron number-density value, $\beta=1$, where it is negative for values of $\beta<1$ (note that experiments such as in laser-solid interactions the population non-degenerate electrons are relatively less than that of degenerate electrons, hence, $\beta\ll 1$) and is positive otherwise. Other such critical $\beta$-values has been reported for quantum two-temperature-electron plasmas \cite{sah}. Also, the coefficients "$B$" vanishes at a critical quantum diffraction parameter value $H_{cr}$ defined as
\begin{equation}\label{Hcr}
H_{cr}=\frac{1}{{1 + \beta }}\sqrt {\frac{\beta }{{1 + \beta }}},
\end{equation}
with a maximum value of $H_m\simeq0.39$ at $\beta=0.5$. This feature is very similar to ones previously reported for ion-acoustic \cite{haas} as well as electrostatic \cite{akbari1} solitary excitations in quantum plasmas. Another similar behavior to mentioned is the independence of $A$-coefficient (i.e. first-order soliton amplitude) from the quantum diffraction parameter, $H$, which is also reported in Refs. \cite{haas, akbari1}.

Figure 1(a) indicates two distinct (dark-bright) regions in $H$-$\beta$ plane, separated by $H_{cr}$ curves (Eq. (\ref{Hcr}), where bright- or dark-type solitary excitations may occur. Also, according to Fig. 1(b), increase in the relative non-degenerate electron density, $\beta$, for fixed  other plasma parameters leads to increase in the soliton amplitude for both rarefactive and compressive solitary excitations. On the other hand, Fig. 1(c) shows that for $H=0.25$ with increase in the value of $\beta$, the soliton width ($B$-coefficient) vanishes at the at some $\beta$-values (defined through Eq. (\ref{Hcr}) but $H=0.5$ it does not. Therefore, for a given fixed value of $H<H_m$, by increase in the value of $\beta$ always one critical $\beta$-value is encountered (Figs. 1(a) and 1(c)) but for $H>H_m$ critical $\beta$-value does not exists. It is important to note that vanishing one of the coefficients $A$ or $B$ (such as in critical values) destroys the KdV evolution equation and no possible KdV-type solitary excitations occur for these special cases, hence, they are excluded from this work. Surprisingly, the shape of solitary excitation is defined by $H_{cr}$, although the soliton amplitude is completely independent of the value of quantum diffraction parameter, $H$. For instance, for a fixed $\beta$-value, increase of the value of the quantum diffraction parameter, $H$, results in different compressive (shown by solid-lines in Fig. 1(d)) and rarefactive (shown by dashed-lines in Fig. 1(d)) branches connected at critical $H$-values. This is much similar to the findings at Ref. \cite{akbari1}.

Figures 1(e) and 1(f) show the dependence of the collision phase-shift on the relative non-degenerate electron population, $\beta$, and the collision angle for different other fixed plasma parameters. It is revealed from Fig. 1(e) that, for the compressive/rarefactive wave branches (solid-curves/dashed curves) the collision phase-shift is negative/positive. Therefore, the sign of the collision phase-shift also can be determined by Fig. 1(a) (Eq. (\ref{Hcr}). Consequently, the in crease in $\beta$-value leads to decrease/increase in the value of collision phase-shift for compressive/rarefactive solitary collisions. It should be noted that, a negative phase-shift in a collision indicates that the collided parts of solitons lag-behind the initial wave trajectory, while, a positive phase-shift means that, the collided parts of solitons move-ahead of the initial wave trajectory \cite{akbari}. Furthermore, it is observed from Fig. 1(f) that for both compressive (solid-curve) and rarefactive (dashed-curve) collisions the absolute-value of phase-shift decreases with increase of the collision angle, $\theta$ but never changes the sign as it has been reported for some plasmas \cite{akbari}.

\section{Concluding Remarks}\label{conclusion}

The extended Poincar\'{e}-Lighthill-Kuo (PLK) reductive perturbation method was used to investigate a 2-dimensional collision of small-amplitude electron-acoustic solitary waves in an unmagnetized two-electron-populated quantum plasma. It is remarked that the propagation and collision properties of these waves is determined by a critical plasma value and significantly dependents on the degeneracy-population. It is also observed that the collision phase-shift is also affected by both relative non-degenerate electron population and the quantum diffraction parameter in such plasmas.

\appendix

\section{Normalized plasma equations in strained coordinate}
\begin{equation}\label{strain1}
\begin{array}{l}
{\varepsilon ^2}\frac{{\partial {n_\alpha }}}{{\partial \tau}} - {c_1}\frac{{\partial {n_\alpha }}}{{\partial \xi}} - {\varepsilon ^2}{c_1}\frac{{\partial {Q_0}}}{{\partial \xi}}\frac{{\partial {n_\alpha }}}{{\partial \eta}} - {c_2}\frac{{\partial {n_\alpha }}}{{\partial \eta}} - {\varepsilon ^2}{c_2}\frac{{\partial {P_0}}}{{\partial \eta}}\frac{{\partial {n_\alpha }}}{{\partial \xi}} +  \\
{k_1}\frac{{\partial {n_\alpha }{u_\alpha }}}{{\partial \xi}} + {\varepsilon ^2}{k_1}\frac{{\partial {Q_0}}}{{\partial \xi}}\frac{{\partial {n_\alpha }{u_\alpha }}}{{\partial \eta}} + {k_2}\frac{{\partial {n_\alpha }{u_\alpha }}}{{\partial \eta}} + {\varepsilon ^2}{k_2}\frac{{\partial {P_0}}}{{\partial \eta}}\frac{{\partial {n_\alpha }{u_\alpha }}}{{\partial \xi}} +  \\
{l_1}\frac{{\partial {n_\alpha }{v_\alpha }}}{{\partial \xi}} + {\varepsilon ^2}{l_1}\frac{{\partial {Q_0}}}{{\partial \xi}}\frac{{\partial {n_\alpha }{v_\alpha }}}{{\partial \eta}} + {l_2}\frac{{\partial {n_\alpha }{v_\alpha }}}{{\partial \eta}} + {\varepsilon ^2}{l_2}\frac{{\partial {P_0}}}{{\partial \eta}}\frac{{\partial {n_\alpha }{v_\alpha }}}{{\partial \xi}} + \ldots  = 0, \\
\end{array}
\end{equation}
\begin{equation}\label{strain2}
\begin{array}{l}
{\varepsilon ^2}{n_\alpha }\frac{{\partial {u_\alpha }}}{{\partial \tau }} - {c_1}{n_\alpha }\frac{{\partial {u_\alpha }}}{{\partial \xi }} - {\varepsilon ^2}{c_1}{n_\alpha }\frac{{\partial {Q_0}}}{{\partial \xi }}\frac{{\partial {u_\alpha }}}{{\partial \eta }} - {c_2}{n_\alpha }\frac{{\partial {u_\alpha }}}{{\partial \eta }} - {\varepsilon ^2}{c_2}{n_\alpha }\frac{{\partial {P_0}}}{{\partial \eta }}\frac{{\partial {u_\alpha }}}{{\partial \xi }} +  \\
{k_1}{n_\alpha }{u_\alpha }\frac{{\partial {u_\alpha }}}{{\partial \xi }} + {\varepsilon ^2}{k_1}{n_\alpha }{u_\alpha }\frac{{\partial {Q_0}}}{{\partial \xi }}\frac{{\partial {u_\alpha }}}{{\partial \eta }} + {k_2}{n_\alpha }{u_\alpha }\frac{{\partial {u_\alpha }}}{{\partial \eta }} + {\varepsilon ^2}{k_2}{n_\alpha }{u_\alpha }\frac{{\partial {P_0}}}{{\partial \eta }}\frac{{\partial {u_\alpha }}}{{\partial \xi }} +  \\
{l_1}{n_\alpha }{v_\alpha }\frac{{\partial {u_\alpha }}}{{\partial \xi }} + {\varepsilon ^2}{l_1}{n_\alpha }{v_\alpha }\frac{{\partial {Q_0}}}{{\partial \xi }}\frac{{\partial {u_\alpha }}}{{\partial \eta }} + {l_2}{n_\alpha }{v_\alpha }\frac{{\partial {u_\alpha }}}{{\partial \eta }} + {\varepsilon ^2}{l_2}{n_\alpha }{v_\alpha }\frac{{\partial {P_0}}}{{\partial \eta }}\frac{{\partial {u_\alpha }}}{{\partial \xi }} -  \\
{k_1}{n_\alpha }\frac{{\partial \varphi }}{{\partial \xi }} - {\varepsilon ^2}{k_1}{n_\alpha }\frac{{\partial {Q_0}}}{{\partial \xi }}\frac{{\partial \varphi }}{{\partial \eta }} - {k_2}{n_\alpha }\frac{{\partial \varphi }}{{\partial \eta }} - {\varepsilon ^2}{k_2}{n_\alpha }\frac{{\partial {P_0}}}{{\partial \eta }}\frac{{\partial \varphi }}{{\partial \xi }} +  \\
\frac{D_{\alpha}}{2}{n_\alpha }{k_1}\frac{{\partial {n_\alpha }}}{{\partial \xi }} + \frac{D_{\alpha}}{2}{\varepsilon ^2}{n_\alpha }{k_1}\frac{{\partial {Q_0}}}{{\partial \xi }}\frac{{\partial {n_\alpha }}}{{\partial \eta }} + \frac{D_{\alpha}}{2}{n_\alpha }{k_2}\frac{{\partial {n_\alpha }}}{{\partial \eta }} +  \\
\frac{D_{\alpha}}{2}{\varepsilon ^2}{n_\alpha }{k_2}\frac{{\partial {P_0}}}{{\partial \eta }}\frac{{\partial {n_\alpha }}}{{\partial \xi }} - {\varepsilon ^2}{k_1}\frac{{{H^2}}}{4}\frac{{{\partial ^3}{n_\alpha }}}{{\partial {\xi ^3}}} - {\varepsilon ^2}{k_2}\frac{{{H^2}}}{4}\frac{{{\partial ^3}{n_\alpha }}}{{\partial {\eta ^3}}} +  \ldots  = 0, \\
\end{array}
\end{equation}
\begin{equation}\label{strain3}
\begin{array}{l}
{\varepsilon ^2}{n_\alpha }\frac{{\partial {v_\alpha }}}{{\partial \tau }} - {c_1}{n_\alpha }\frac{{\partial {v_\alpha }}}{{\partial \xi }} - {\varepsilon ^2}{c_1}{n_\alpha }\frac{{\partial {Q_0}}}{{\partial \xi }}\frac{{\partial {v_\alpha }}}{{\partial \eta }} - {c_2}{n_\alpha }\frac{{\partial {v_\alpha }}}{{\partial \eta }} - {\varepsilon ^2}{c_2}{n_\alpha }\frac{{\partial {P_0}}}{{\partial \eta }}\frac{{\partial {v_\alpha }}}{{\partial \xi }} +  \\
{k_1}{n_\alpha }{u_\alpha }\frac{{\partial {v_\alpha }}}{{\partial \xi }} + {\varepsilon ^2}{k_1}{n_\alpha }{u_\alpha }\frac{{\partial {Q_0}}}{{\partial \xi }}\frac{{\partial {v_\alpha }}}{{\partial \eta }} + {k_2}{n_\alpha }{u_\alpha }\frac{{\partial {v_\alpha }}}{{\partial \eta }} + {\varepsilon ^2}{k_2}{n_\alpha }{u_\alpha }\frac{{\partial {P_0}}}{{\partial \eta }}\frac{{\partial {v_\alpha }}}{{\partial \xi }} +  \\
{l_1}{n_\alpha }{v_\alpha }\frac{{\partial {v_\alpha }}}{{\partial \xi }} + {\varepsilon ^2}{l_1}{n_\alpha }{v_\alpha }\frac{{\partial {Q_0}}}{{\partial \xi }}\frac{{\partial {v_\alpha }}}{{\partial \eta }} + {l_2}{n_\alpha }{v_\alpha }\frac{{\partial {v_\alpha }}}{{\partial \eta }} + {\varepsilon ^2}{l_2}{n_\alpha }{v_\alpha }\frac{{\partial {P_0}}}{{\partial \eta }}\frac{{\partial {v_\alpha }}}{{\partial \xi }} -  \\
{l_1}{n_\alpha }\frac{{\partial \varphi }}{{\partial \xi }} - {\varepsilon ^2}{l_1}{n_\alpha }\frac{{\partial {Q_0}}}{{\partial \xi }}\frac{{\partial \varphi }}{{\partial \eta }} - {l_2}{n_\alpha }\frac{{\partial \varphi }}{{\partial \eta }} - {\varepsilon ^2}{l_2}{n_\alpha }\frac{{\partial {P_0}}}{{\partial \eta }}\frac{{\partial \varphi }}{{\partial \xi }} +  \\
\frac{D_{\alpha}}{2}{n_\alpha }{l_1}\frac{{\partial {n_\alpha }}}{{\partial \xi }} + \frac{D_{\alpha}}{2}{\varepsilon ^2}{n_\alpha }{l_1}\frac{{\partial {Q_0}}}{{\partial \xi }}\frac{{\partial {n_\alpha }}}{{\partial \eta }} + \frac{D_{\alpha}}{2}{n_\alpha }{l_2}\frac{{\partial {n_\alpha }}}{{\partial \eta }} +  \\
\frac{D_{\alpha}}{2}{\varepsilon ^2}{n_\alpha }{l_2}\frac{{\partial {P_0}}}{{\partial \eta }}\frac{{\partial {n_\alpha }}}{{\partial \xi }} - {\varepsilon ^2}{l_1}\frac{{{H^2}}}{4}\frac{{{\partial ^3}{n_\alpha }}}{{\partial {\xi ^3}}} - {\varepsilon ^2}{l_2}\frac{{{H^2}}}{4}\frac{{{\partial ^3}{n_\alpha }}}{{\partial {\eta ^3}}} +  \ldots  = 0, \\
\end{array}
\end{equation}
\begin{equation}\label{strain4}
\begin{array}{l}
{\varepsilon ^2}\lambda_1^2\frac{{{\partial ^2}\varphi }}{{\partial {\xi^2}}} + {\varepsilon ^2}\lambda_2^2\frac{{{\partial ^2}\varphi }}{{\partial {\eta^2}}} + 2{\varepsilon ^2}\mu\frac{{{\partial ^2}\varphi }}{{\partial \xi\partial \eta}}  - \sum\limits_\alpha  {{n_\alpha }} +
z_i n_i + \ldots =0.
\end{array}
\end{equation}

\newpage

\textbf{FIGURE CAPTIONS}

\bigskip

Figure-1

\bigskip

(Color online) Figure 1(a) shows the regions dark-bright in $\beta$-$H$ plane for which the solitary excitations are compressive (grey-regions) or rarefactive (bright-regions) and accordingly where the collision phase-shifts are positive (bright-regions) or negative (grey-regions). Figure 1(b), 1(c) and 1(d) show the variations of soliton amplitude and width with respect to the fractional non-degenerate electron population $\beta$ and quantum diffraction parameter, $H$ for different values of other plasma parameters (critical $\beta$- and $H$-values are shown in Figs. 1(c) and 1(d)). The thick curve in plot 1(c) corresponds to the value of $H=0.5$. Figure 1(e) and 1(f) depicts the variation of collision phase-shift with respect to fractional non-degenerate electron population $\beta$ and the collision angle, $\theta$ for compressive (solid-curves) and rarefactive (dashed-curve) in oblique solitary interactions. The values of $\varepsilon=0.1$ and $\lambda_1=\lambda_2=0.1$ are used in all plots. Different dash sizes in plots 1(d) and 1(e) are used to appropriately presents different values of varied parameter in each plot.

\newpage

\bigskip
\begin{figure}[ptb]\label{Figure1}
\includegraphics[scale=.5]{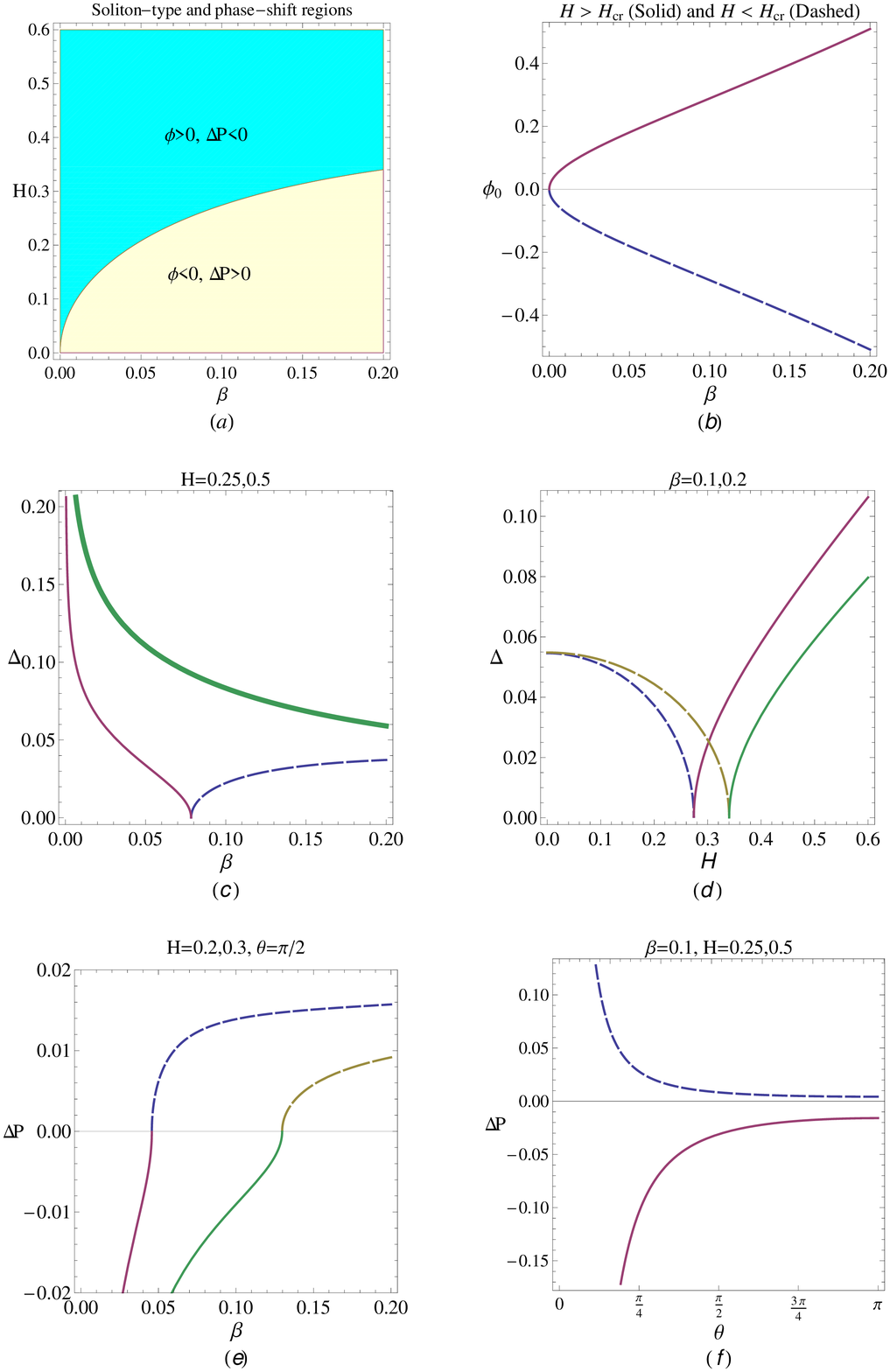}\caption{}
\end{figure}

\end{document}